# LLM-assisted sentiment analysis for integrated computational and qualitative mixed methods education research: A case study of students' written reflection assignments

Xiomara Gonzalez[1], Gabriella Coloyan Fleming[2], Andrew Katz[2], Maya Denton[3], Jessica Deters[4]

[1]Chandra Family Department of Electrical and Computer Engineering, University of Texas at Austin
[2]Department of Engineering Education, Virginia Polytechnic Institute and State University
[3]Gallogly College of Engineering, University of Oklahoma
[4]Department of Mechanical and Materials Engineering, University of Nebraska-Lincoln

**Abstract**

Written reflection assignments give students valuable opportunities for critical self-assessment, meaning making, and learning processing. Additionally, such reflections provide rich data for qualitative education research. However, qualitative data can be time-consuming to analyze. It is even more time-intensive to qualitatively compare findings between different groups of participants, usually limiting comparison to, at most, one variable (e.g., binary gender). Large language models (LLMs) have recently begun to be critically evaluated for use as qualitative research assistants. Using a longitudinal case of written student reflections ($n$ = 151) from a study abroad program, we investigate how LLM-assisted sentiment analysis can enable longitudinal mixed-methods research combining computational and thematic analyses. First, statistical testing is used to quantitatively compare sentiment differences according to seven different student identity/lived experience variables. Then, these results inform qualitative data analysis to investigate the reasons underlying these differences. For the case of undergraduate students studying abroad, we found that prior experience living abroad was the only personal variable impacting students' sentiments of their verbal language and communication behaviors. This workflow has implications for how qualitative researchers can more easily probe multiple variables when comparing participants from different demographic groups.

## 1 | Introduction

Cross-cultural communication is a crucial skill in the global workforce, critical for trade, business management, marketing, negotiation, interpersonal relationships, and teambuilding (Jenifer & Raman, 2015; Padhi, 2016). As such, it is key that undergraduate programs prepare students for success in the global workforce. Cross-cultural communication skills are particularly important in computing and engineering because of the international nature of engineering design, development, and manufacturing (Bremer, 2008; Parkinson, 2009). Study abroad opportunities have emerged as an effective strategy to develop students' cross-cultural communication (Davis & Knight, 2018).

Written reflections are an active learning pedagogical tool used to critically engage students through meaning-making and processing learning (Chan & Wong, 2023; Turns et al., 2014). Reflection assignments have been implemented in many non-study abroad undergraduate engineering courses, such as reflective comments on technical lab reports (Selwyn & and Renaud-Assemat, 2020), ePortfolios in project-based curricula (Demetry et al., 2019), structured written reflections in first-year engineering courses (Diefes-Dux & Cruz Castro, 2022), and "exam wrappers" after students receive graded exams (Gamieldien et al., 2023). Written reflections or journaling assignments have also become a common practice in engineering study abroad programs (Fleming et al., 2024; Gough et al., 2018; Knight et al., 2019; Morgan et al., 2021) because they provide students with structure and space to reflexively process their experiences (Deters et al., 2022; Glass, 2014). Engineering study abroad courses use reflective writing as purposive exercise to help students make sense of and express their experiences (Berka et al., 2021; Deters et al., 2022; Glass, 2014; Jaiswal et al., 2024).

This study has two goals. First, we seek to understand how LLMs can be used to aid researchers in conducting longitudinal computational and qualitative analysis. Second, we apply this mixed-method approach to investigate undergraduate students' sentiments in reflections from an engineering study abroad course.

This study was guided by two research questions:

- RQ1: How can a mixed-methods approach of LLM-assisted sentiment analysis followed by thematic analysis aid education researchers in analyzing qualitative data?
- RQ2: For the case of an engineering study abroad program, how do undergraduate students' attitudes about their verbal language and communication differ based on the students' identities and lived experiences?

## 2 | Literature review

### 2.1 | Sentiment analysis

One benefit of traditional qualitative data analysis is the ability to identify expressed sentiments and interpret them within their contextual nuances. The automated counterpart to this process is comparable to sentiment analysis, a subfield of natural language processing (NLP), which analyzes people's opinions, sentiments, evaluations, appraisals, attitudes, and emotions towards entities, such as products, events, and topics (Liu, 2022). Sentiment analysis is commonly framed as a multiclass classification task in which the polarity of sentiment expressed in text is labeled as positive, negative, or neutral (Nasukawa & Yi, 2003). Sentiment analysis approaches vary across disciplines. For example, machine learning (ML)-based sentiment analysis has included both supervised and unsupervised methods such as training models on labeled sentiment data or clustering data to identify sentiment patterns (Tan et al., 2023; Wankhade et al., 2022). More recently, neural network transformer architectures that serve as the foundation for LLMs have been shown to rival established methods for sentiment classification (Krugmann & Hartmann, 2024).

Prior education-focused sentiment analysis studies have explored the performance of ML-based models to assess student reflections on their course experiences (Baisley & Marutla, 2024; Lazrig & Humpherys, 2022; Roy & Rambo-Hernandez, 2021). More recently, researchers have begun to use LLMs to conduct sentiment analysis. Within education, LLM-assisted sentiment analysis has most commonly been used to examine student evaluations of teaching, in which students provide feedback to their instructors at the conclusion of a semester (Lacy et al., 2026; Peña-Torres, 2024; Shaikh et al., 2023). LLMs have also been used to analyze written student reflections on what they found interesting and confusing during learning (Satya Putra et al., 2025). Researchers train the LLM on a subset of the data before using it to analyze the full dataset (Satya Putra et al., 2025; Shaikh et al., 2023). Cohen's kappa, a metric historically used to compare the inter-rater reliability between two human coders Landis and Koch's (Landis & Koch, 1977), has been adopted to calculate inter-rater agreement between a human coder and an LLM. In these studies, Cohen's kappa values were 0.58 (moderate agreement) for GPT-3.5 (Shaikh et al., 2023) and 0.588 (moderate agreement) for GPT-4 (Satya Putra et al., 2025).

However, there are several caveats for LLM-assisted sentiment analysis. LLMs are less effective when complex sentiment analysis requires a deeper understanding of sentiment phenomena or structure, compared to handling simpler classifications (Zhang et al., 2024). Additionally, models can vary greatly, leading to reduced reproducibility, bias amplification, and unstable sentiment classification (Herrera-Poyatos et al., 2025). Despite these limitations, the classification of sentiment into positive, negative, and neutral is a well-scoped task for LLMs to complete, and the moderate inter-rater agreement observed in prior studies suggests this approach can yield results consistent with human judgement.

## 2.2 | Computational and qualitative "mixed methods" research

Sentiment analysis provides a way to quantitatively analyze large corpora of qualitative data. Accordingly, researchers have combined computational and qualitative methods into a new type of mixed methods research (Al-Garaady & Albuhairy, 2025; Andreotta et al., 2019; Guenduez et al., 2025; Love et al., 2018; Wanniarachchi et al., 2023). As presented in Table 1, the quantitative methods largely use NLP, and qualitative methods use human-based analysis. In all these instances, quantitative methods preceded the qualitative methods. Sentiment analysis is the most common quantitative method; thematic analysis is the most common qualitative method. Only one approach used NLP, rather than a human coder, to analyze the qualitative data (Wanniarachchi et al., 2023).

**Table 1. Summary of computational and qualitative "mixed methods" approaches**

| Publication | Quantitative Method | Qualitative Method(s) |
|---|---|---|
| Albuhairy & Algaraady, 2025 | Sentiment analysis (NLP) | Thematic analysis (human) |
| Andreotta et al., 2019 | Topic alignment (NLP) | Thematic analysis (human) |
| Guenduez et al., 2025 | Structural topic modeling (NLP) | Interpretation of topics (human; method unspecified) |
| Love et al., 2018 | Count-based sentiment analysis for specified words (human and computer) | Abductive coding (human) |
| Wanniarachchi et al., 2023 | Sentiment analysis (NLP) | Topic modeling, discourse analysis (NLP) |

## 2.3 | Large language models and qualitative analysis

Qualitative data analysis (QDA) has traditionally been a labor-intensive process for researchers, but publicly available LLMs have revolutionized QDA by increasing both the speed of analysis and size of data sets (De Morais Leça et al., 2025). Closed source generative large LLMs (sometimes called "commercial models"), such as OpenAI's Chat Generative Pre-Trained Transformer (ChatGPT), have been explored to support deductive coding (Katz et al., 2023; Tai et al., 2024; Xiao et al., 2023), inductive thematic analysis (Gamieldien et al., 2023; Jaiswal et al., 2024; Katz et al., 2023), and the development of tools for collaborative coding (Gao et al., 2024). Outputs from LLMs are also often treated as an additional individual coder (Schiavone et al., 2023; Xiao et al., 2023). However, one of the major drawbacks of closed source models, especially when working with sensitive participant data, is that they rely on external services for data processing. Data sent through such platforms is often retained by LLM developers. The lack of transparency regarding how input data is repurposed and contributes to future LLM training exacerbates privacy, ethical, and security concerns (Alder, 2023). Another potential pitfall for researchers using closed source models is that these models are typically accessed through software intermediaries, without the option to locally store model versions. The duration of the public's access to model variations is at the discretion of the providing entity, meaning research approaches cannot always be reproduced consistently. Open source generative LLMs are comparable to their closed source counterparts, offering similar performance on a variety of benchmarks, including multi-choice question answering on a range of academic subjects and human level common sense inference (e.g., sentence completion) (Artificial Analysis, 2026). Researchers can store and run these LLMs on local hardware, thus mitigating concerns about data processing. The open source community has also developed LLM variations that can operate on a range of computational platforms, including on personal laptops, offering the ethical benefit of protecting participants' data. This development is particularly important for protected health information (Mathis et al., 2024) or when participants only consented for their data to be shared with the researchers.

In computing education research, LLMs have assisted human researchers in content analysis of two decades of computing education research (Gale & Nicolajsen, 2025) and qualitative analysis of K-12 educators' use of AI (A. Liu et al., 2025). More broadly, education researchers have conducted LLMs-assisted qualitative analysis on myriad topics, including math tutoring lessons (Barany et al., 2024; X. Liu et al., 2024), student essays about their career interests (Katz et al., 2023), teaching evaluations (Katz et al., 2024), students' understanding of engineering mechanics and thermodynamics concepts (Auby et al., 2025), and computer science attrition (Ross & Katz, 2025).

Although many researchers seek to use LLMs for automation, an alternative approach is to leverage them as collaborative research partners. In Sinha et al.'s (2024) workflow, two human researchers first conducted open coding to independently produce two codebooks, which they then discussed to reach consensus in the form of a collaborative human-generated code book. Separately, GPT-4 was used to create a codebook. The two human researchers then compared and discussed the human- and GPT-4-generated codebooks, incorporating findings from both to create the codebook. Meng et al.'s (2025) CHALET (**C**ollaborative **H**uman-LLM **A**na**L**ysis for Empowering **T**heory-driven qualitative coding) approach is similar, using an LLM-generated codebook to inform deeper human analysis. Human researchers and gpt-4-1106-preview deductively coded the data in parallel, and then the human researchers conducted a disagreement analysis comparing the human- and LLM-generated findings. This process was found to produce advanced insights that were not found by human researchers alone.

**2.4 | Cross-cultural communication**

Studies have shown that studying abroad provides engineering students with opportunities to develop and improve cross-cultural communication skills. In some cases, students report cross-cultural communication as their foremost learning outcome from studying abroad, such as the abilities to convey one's desired message and to recognize if teammates understood the message (Davis & Knight, 2018). When study abroad programs include work experience (i.e., internships or research), acquired verbal language skills include gaining and using technical vocabulary in a foreign language (Berka et al., 2021). In addition to specific aspects of cross-cultural communication, study abroad also teaches students about the importance of intercultural skills for effective communication, particularly for engineering work (Jaiswal et al., 2024; Wrobetz et al., 2024). U.S. students studying abroad in non-English-speaking or culturally dissimilar countries are more likely to mention gaining cross-cultural communication skills (Knight et al., 2019, 2020). Incidents around communication help students connect engineering and cultural learning (Jaiswal et al., 2024) and learn to navigate a foreign country (Davis & Knight, 2025). Recognizing the importance of studying abroad for intercultural communication development, some engineering study abroad programs incorporate communication-related course activities and learning outcomes (Berger & Bailey, 2013; Davis & Knight, 2018).

Longitudinal qualitative analyses of student reflections show that different aspects of the study abroad experience are salient to students at different points throughout their programs. Communication is a key focus early on when students are studying in countries where the language is not their native tongue (Savicki et al., 2007; Wrobetz et al., 2024). Over time, students adapt to understanding others in their host country and making themselves understood in a variety of settings, including their living situations, immediate environments, meeting places, and in public, focusing both on interpersonal communication patterns and expectations. By the end of their programs, students feel they are more able to communicate with people from other cultures (Savicki et al., 2007) and are more aware that effective communication is an important skill for engineering work (Wrobetz et al., 2024). Longitudinal analysis of engineering students' reflections has shown how they develop different global engineering competencies (i.e., technical, professional, and global) over the course of their study abroad programs, and that this longitudinal competence development is highly context-specific (see Schuman, 2025 for more detail).

**3 | Current Study**

The current study furthers existing research on computational and qualitative mixed-methods research, LLMs as collaborative researchers, and undergraduate students' cross-cultural communication skills. First, we seek to extend computational and qualitative methods from NLP-human workflows to an LLM-human workflow (Andreotta et al., 2019; Guenduez et al., 2025). Traditional mixed methods utilize two distinct sets of quantitative and qualitative data. Here, our "quasi-mixed" method applies quantitative and qualitative analysis to the same qualitative dataset. Figure 1 shows two traditional forms of mixed-methods research: convergent parallel design (Fig. 1a) and explanatory sequential design (Fig. 1b) (Creswell, 2014); Fig. 1c shows this study's design. Secondly, we follow Sinha et al. (2024) and Meng et al. (2025) in utilizing an LLM as a collaborative coding partner. Finally, we apply this integrated computational and qualitative approach to a case study, longitudinally examining undergraduate students' sentiments of their verbal language and communication behaviors over the course of an engineering study abroad course.

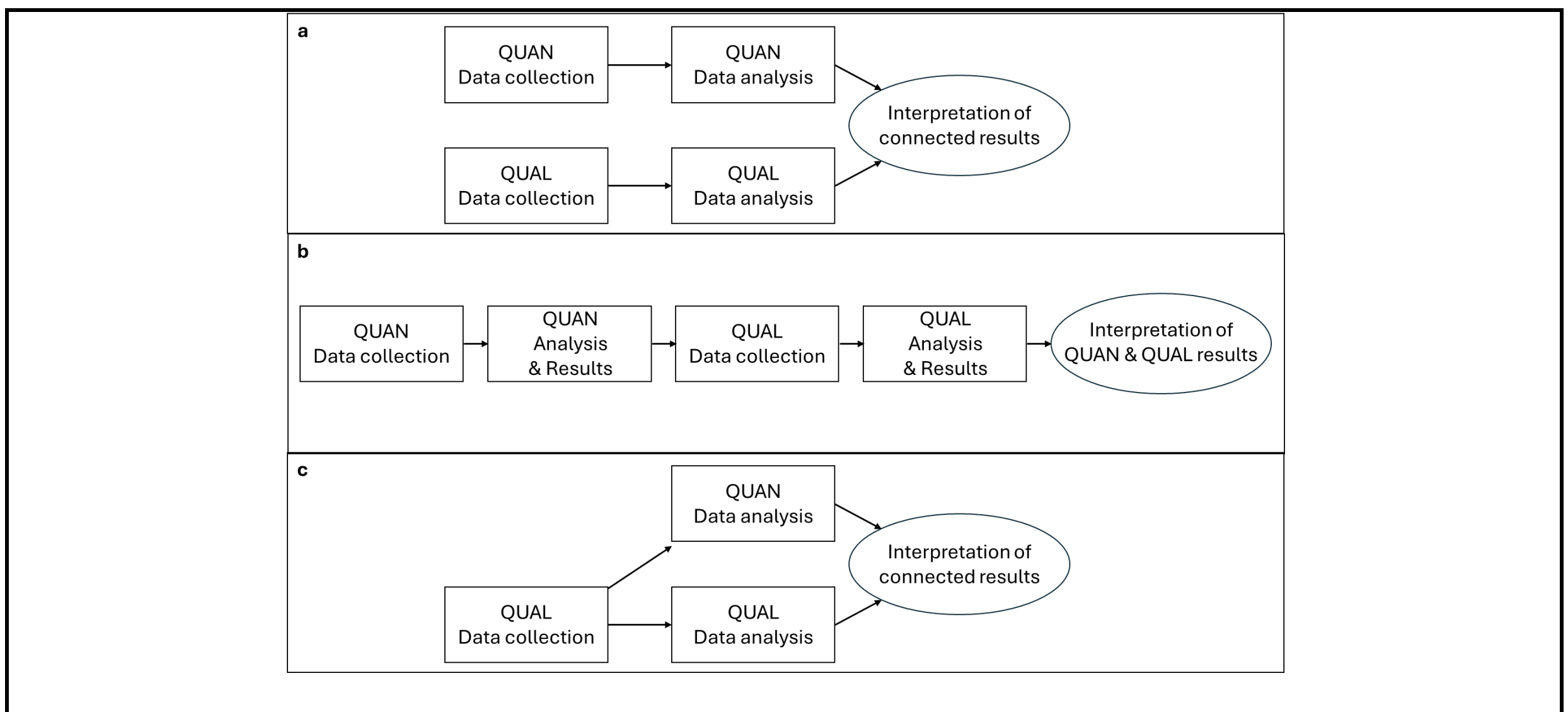

**Figure 1. (a) Convergent parallel design, (b) explanatory sequential design, (Creswell, 2014), (c) this study's quasi-convergent parallel design**

## 4 | Methods

### 4.1 | Overview of the study abroad program

This study's dataset is a subset from a larger study incorporating written reflections in a month-long study abroad course (MASKED FOR REVIEW). The course taught was a U.S. institution's offering of a three-credit, undergraduate-level engineering elective open to students of all majors. Similar to other study abroad programs, the program condensed a semester-long course into a month-long experience to mitigate barriers to studying abroad such as course credit transfers, financial resources, and homesickness (Besser et al., 2017; Ferreira et al., 2020). Eighty students, three graduate teaching assistants, three faculty, and one researcher-instructor from the institution traveled together to Japan. Before departure, cultural differences between the U.S. and Japan were discussed. On the first day of instruction, students received a brief orientation that included helpful Japanese phrases, such as greetings. The course consisted of lectures, a month-long scaffolded group project, and a five-day field trip to Kyoto and Hiroshima. Prior to the final reflection due date, the researcher-instructor led an in-class discussion about the reflections.

**4.2 | Data collection**

Data in this study are part of students' reflections written throughout the study abroad course. Four two-page reflections were assigned: one at the beginning, two during, and one at the end. A subset of 151 reflections from three of the four time points was used for our analysis. Student reflections collected at the second time point were not considered because they did not have questions explicitly about language and communication. For each reflection, only responses to questions prompting students to discuss communication, such as students' comfort levels communicating in their non-primary language, expectations for communicating in Japan, and recalling interactions with locals, were extracted for analysis (Table 2). The full list of reflection questions is available in (MASKED FOR REVIEW).

**Table 2. Reflection questions used for analysis.**

| **Time point** | **Number reflections (N=51)** | **Questions** |
|---|---|---|
| Reflection 1 | 51 | 1. How are you feeling about the trip?<br>a. What languages do you have experience with, and how do you anticipate that experience will impact your time in Japan?<br>2. What would you say is your primary culture?<br>a. What are the characteristics of this culture?<br>b. What is this culture's spoken/written language? Are there any characteristic aspects of non-verbal communication?<br>3. Have you ever experienced another culture before (e.g., through travel, family)?<br>a. What are the characteristics of this culture?<br>b. What is this culture's spoken/written language? Are there any characteristic aspects of non-verbal communication? |

| Reflection 3 | 49 | 1. Reflect on how you have communicated with locals so far on this trip.<br>  a. Do you have experience with a language other than English?<br>  b. Do you have any experience with the Japanese language?<br>2. Reflect on your expectations with respect to the language difference in Japan.<br>  a. What have you noticed about non-verbal communication in Japan? |
|---|---|---|
| Reflection 4 | 51 | 1. Read through reflections 1-3. Do you feel more or less comfortable being able to communicate (verbally or non-verbally) in another country where the language is not your primary language? |

**4.3 | Participants**

Out of 80 students, 51 (64%) consented to participate. Of these, the majority (*n*=30, 61%) identified as men. Eighteen (35%) students identified as Asian or Asian American, 15 (21%) as White, 11 (22%) as Hispanic or Latino/a/é, and one (2%) as Middle Eastern or North African. Six (12%) students identified with multiple ethnicities. Most students (*n*=32, 63%) were familiar with one or more languages other than English. Foreign languages known by participants included: Arabic (1 student), Cantonese (1), Chinese (4), French (5), Hindi (3), Japanese (5), Latin (1), Malayalam (1), Mandarin (4), Spanish (12), and Vietnamese (2). The majority (*n*=26, 51%) also had at least one parent born outside of the U.S. Twelve students (24%) identified as first-generation American, and 10 students (20%) are first-generation college attendees. Thirteen students (26%) reported having lived abroad, and 14 (28%) students identified as low-income. The majority (*n*=43, 84%) were engineering majors from across the College of Engineering. We do not provide intersectional identities for participant privacy.

At the end of the course, students could opt-in to have their reflections included in the study. All students chose to include the first and final reflections. Two students omitted Reflections 3. All participants were provided with a $25 Amazon gift card to thank them for their participation.

### 4.4 | Data analysis

Initial analysis through analytic memos and discussions found that students had many negative perceptions of their communication skills. Figure 2 outlines our six-stage data analysis pipeline: (1) processing student reflections and data familiarization, (2) manual quote extraction, (3) human researcher and LLM-assisted sentiment analysis of student quotes mentioning verbal language and communication behaviors, (4) human researcher's revision of LLM sentiment labels, (5) qualitative and quantitative analyses, and (6) interpretation of qualitative findings considering quantitative results. Denoted in red are participants' data, green are tasks completed by the human researcher, and blue are tasks utilizing the LLM.

#### 4.4.1 | Data processing, manual quote extraction, and sentiment labeling

In Stage 1, the first author read all reflections and wrote analytic memos for each reflection time point to familiarize themselves with the data. In Stage 2, the first author isolated and extracted student responses to questions on their language and communication behaviors (Table 2). Verbal language and communication behaviors were defined as spoken language and communication skills and abilities (Van Dyne et al., 2012). This included instances where students adjusted their accent, tone, speaking rate, and style of expression when talking. We focused on verbal behaviors as students referenced this aspect more than nonverbal or written communication. In Stage 3, the first author assigned each quote an overall sentiment label of positive, negative, or neutral (Baisley & Marutla, 2024; Lazrig & Humpherys, 2022), and consulted with the research team throughout the process to discuss instances where sentiment was nuanced to ensure consistent labeling.

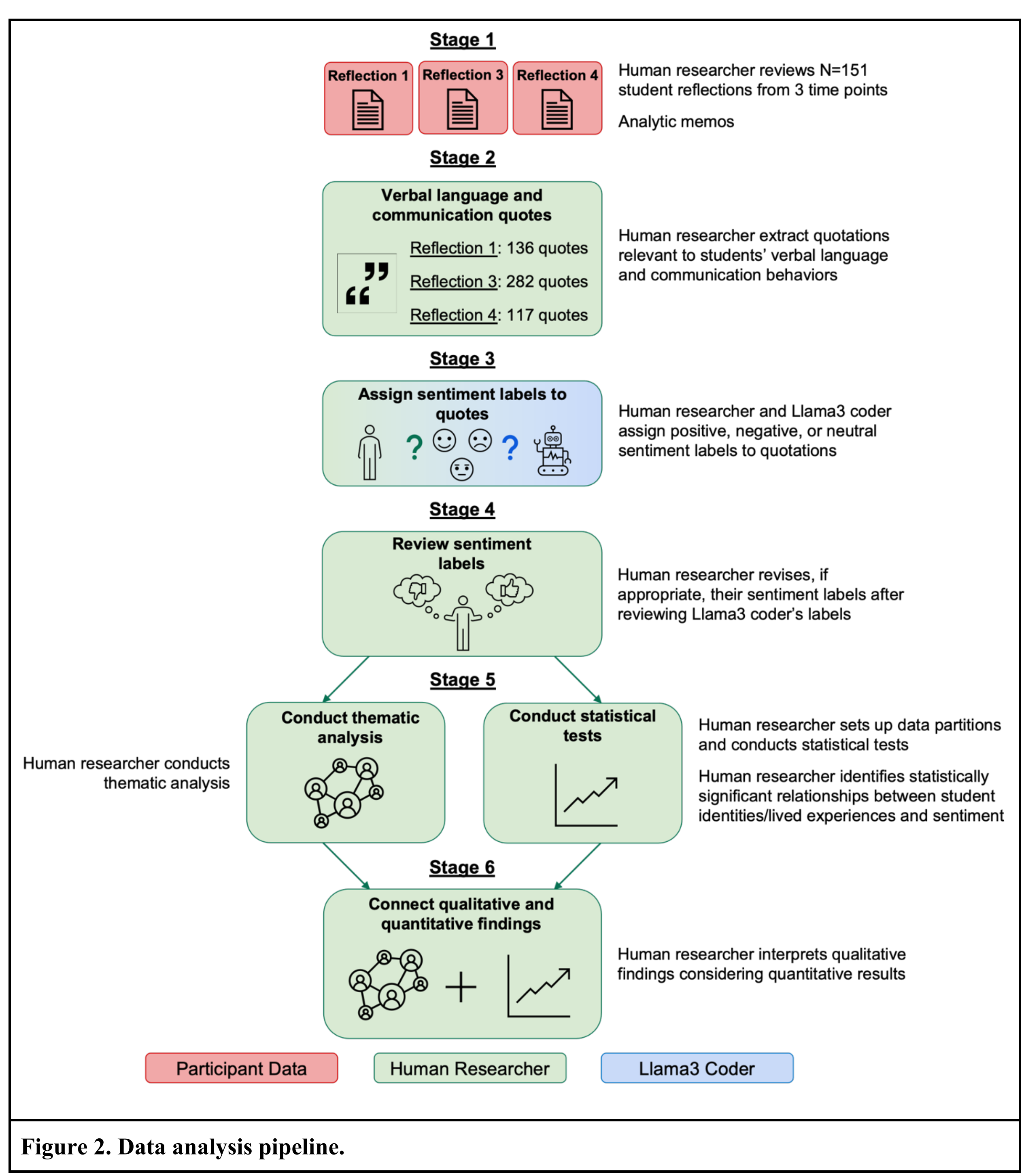

**Figure 2. Data analysis pipeline.**

### 4.4.2 | Large language model-driven sentiment analysis of quotes

When this paper was written, to the best of our knowledge, there were only a limited number of studies evaluating smaller open-source models (with fewer than 10 billion parameters) for data analysis tasks, such as classification and text extraction from unstructured text summaries (Perron et al., 2024). In Stage 3, we used MetaAI's open source Llama3 LLM model (8 billion parameters) to conduct sentiment analysis. We used the Ollama framework to download and test open source models from different families, including MetaAI's Llama2 and Llama3, Mistral AI's Mistral-Nemo, and Alibaba Cloud's Qwen, across three model size brackets, ranging from ~7 billion to 70 billion parameters. MetaAI's Llama3 model (8 billion parameters) was ultimately selected as it was the latest version at the time of data analysis. It delivered the highest interrater reliability and provided several balanced tradeoffs. First, smaller models (<10 billion parameters) minimize computational resources required for task execution, as they can be run on personal computers, facilitating the adoption of our study methods in other contexts. Small models also take less time to process a prompt. Second, Llama3's model hyperparameters–such as temperature, which ranges from 0 to 1 and regulates the randomness of the model's responses (MetaAI, 2024)–can be modified as a control for how much the model deviates from prompt instructions. Third, student reflections may disclose sensitive information and are subject to Institutional Review Board policies and procedures protecting participants. For example, federal and higher education institutions have established acceptable use policies for generative AI use in research activities by defining risk levels for institutionally-owned data and restrictions on which tools can be used (U.S. Department of Health and Human Services Office of Human Research Protections, 2022). Our participants did not consent for their data to be shared with third-party AI-driven commercial entities, only members of the research team. Thus, to ethically conduct this study, we only considered open source models.

Llama3 was prompted to provide a sentiment label of positive, negative, or neutral for all student quotes that were manually extracted. Appendix A1 outlines seven characteristics used to format our prompt. We approached sentiment labeling as a classification task, using a few-shot prompting method to guide the model by including a few demonstrations of the task in the prompt (Brown et al., 2020). The default values for the Llama3 model inference parameters were used with the exception of temperature, which was set to 0 to produce deterministic, replicable outputs. All model prompting experiments were conducted using a server equipped with eight AMD Radeon Instinct MI50 GPUs, each with 32GB of memory.

In Stage 4, Llama3 outputs were analyzed by the first author to identify the model's sentiment label for individual quotes. There is currently no standardized method as to how an LLM-based output should be compared to a human researcher's data interpretations. Our analysis in Stage 4 adopted a similar approach to Meng et al. (2025) and Sinha et al. (2024): model outputs were treated as an additional individual rater and discrepancies between the human and LLM raters informed deeper analysis. To resolve discrepancies in sentiment labels between the first author and Llama3 model, the first author reexamined the quotation in question, consulting with the research team when sentiment was difficult to determine. If determined that the first author's initial interpretation was inaccurate, the first author revised their label assigned to the quote. Llama3 was not re-prompted to reevaluate its initial output.
Cohen's kappa ($\kappa$) was used to determine the interrater reliability between the first author and Llama3's output. We used Landis and Koch's (1977) nomenclature for describing the relative strength of agreement, following the scale: poor ($\kappa < 0.00$), slight ($0.00 < \kappa < 0.20$), fair ($0.21 < \kappa < 0.40$), moderate ($0.41 < \kappa < 0.60$), substantial ($0.61 < \kappa < 0.80$), and almost perfect ($0.81 < \kappa < 1.0$). Similar to Zambrano et al. (2023), prompts were iterated to increase Cohen's kappa.

#### 4.4.3 | Qualitative analyses of student responses to language and communication questions

In Stage 5, we conducted two rounds of inductive coding on the corresponding student quotations using open coding techniques (Corbin & Strauss, 1998). In the first round, four major themes were identified: 1) students' prior language and communication experiences, 2) preparation for interactions with Japanese locals, 3) language dynamics and their role in verbal communication, and 4) perceived outcomes of study abroad. During the second round of coding, sub-themes were identified and categorized based on how each participant framed their experiences: positive (beneficial to their experience), negative (detrimental), or neutral. The first author completed both rounds of coding. The first, second, and third authors met weekly to discuss the results.

#### 4.4.4 | Quantitative analyses of student responses to verbal language and communication questions

In Stage 5, the first author also conducted quantitative analyses using their revised labels. The dataset of student quotations was partitioned using the following variables: (1) sentiment label: positive, negative, neutral; (2) reflection time point: reflection 1, reflection 3, reflection 4, all reflections; and (3) student identities and lived experiences. Student identities and lived experiences were binary (whether or not the student identified with the characteristic). These included whether students knew one or more foreign languages; had one or more parents born outside of the US; had previously lived abroad; and/or identified as first-generation American, first-generation college attendee, low income and/or Asian/Asian American. These characteristics were chosen because they can be facilitators and barriers to study abroad (Petzold & Moog, 2018; Simon & Ainsworth, 2012). We specifically considered Asian/Asian American—rather than broader categories of race and ethnicity—as the study abroad program took place in Japan, and prior research suggests that study abroad experiences are influenced by students' cultural distance to the host country (Davis & Knight, 2021). For example, a data partition could include individual students' counts of positive sentiments expressed in quotations from Reflection 1, grouped by students who identify as Asian/Asian American and those who do not.

The Shapiro-Wilk test was used to determine whether sentiment count data was normally distributed, with $p$ values $<.05$ indicating statistical significance. Checking for normality ensures that accurate and reliable conclusions are drawn, especially for small sample sizes (Ghasemi & Zahediasl, 2012). For normally distributed data, the parametric Student's *t*-test was conducted; else, the Wilcoxon rank-sum test (Mann-Whitney U test) was used as a non-parametric alternative. We considered *p* values <.05 statistically significant for both parametric and non-parametric tests. Statistical tests identified key data partitions for subsequent qualitative analyses.

Depending on whether a parametric or non-parametric test was conducted, Cohen's *d* or the Glass rank biserial correlation coefficient ($r_g$) was calculated to estimate the strength of the association between students' characteristics and lived experiences and students' sentiments, respectively (King et al., 2018). For Cohen's *d*, the strengths of associations were small ($d = 0.20$), medium ($d = 0.50$), and large ($d = 0.80$) (Cohen, 1988, 1992). For the Glass biserial correlation coefficient, the strengths of the associations were small ($r_g = 0.10$), medium ($r_g = 0.30$), and large ($r_g = 0.50$) (Cohen, 1988, 1992).

In Stage 6, qualitative findings were interpreted in the context of quantitative results. Thematic similarities and differences were examined for statistically significant relationships between student identities and lived experiences and sentiments.

### 4.5 | Limitations

Limitations in our study design and findings that can guide future research. First, our data is drawn from students attending a single large, research-intensive institution in the southwestern United States. A primary focus of our larger study was to have students reflect on many topics, including their language and communication skills and behavior. Although students were encouraged to address all parts of the reflection prompts, some may have only fully responded to a subset of questions. Additionally, students might have shared insights about their verbal language and communication behaviors outside of the specific prompts we focused on in our analysis. Some students arrived in Japan before the program began, and on non-class days, some explored regions with limited English-speaking infrastructure. These variations likely influenced the content of students' reflections. We did not collect data on certain student identities and lived experiences (e.g., prior travel, which could have guided their preparation for the study abroad program).

Second, the measured effect for the quantitative statistical tests for sentiment counts ranges from small to medium, limiting the generalizability of our findings. Our results may not extend to other study abroad programs, particularly those focused on language immersion and proficiency. This limitation also applies to longer study abroad programs, which may offer students more opportunities to interact with native speakers (Meara, 1994).

Third, although our study focuses on students' sentiments toward verbal language and communication behaviors, we acknowledge that nonverbal communication and written language are also key elements shaping students' communication experiences abroad, especially in countries like Japan, where written language holds deep cultural significance (Gottlieb, 2005). Limitations regarding the incorporation of an LLM model in our data analysis pipeline are discussed in section 5.1.2.

## 5 | Results

### 5.1 | RQ1: Mixed-methods approach

#### 5.1.1 | Advantages: Efficiency and analytic support in qualitative analysis

To address RQ1, we examined how an LLM-assisted sentiment analysis pipeline, paired with human-led thematic analysis, can support education researchers in analyzing large volumes of qualitative data. Our findings suggest that LLMs can serve as an effective analytical support tool, particularly in the early stages of coding, by accelerating sentiment labeling while maintaining alignment with human interpretation. Beyond efficiency, LLMs also functioned as generative analytic collaborators. Instances of divergence between LLM- and human-assigned sentiment prompted reflexive discussion, clarification of coding criteria, and closer examination of ambiguous excerpts.

Notably, these divergences were not random. The LLM demonstrated a tendency to assign affective labels (e.g., positive or negative) to statements the human coder interpreted as neutral or descriptive in nature. For example, in response to the student statement "I have experience in Tagalog and Spanish (Reflection 3, Andrew)," the model produced a positive label emphasizing language proficiency, "POSITIVE: student has experience with Tagalog and Spanish and can say specific phrases well." In contrast, the human coder coded this excerpt as neutral, as it reflects prior experience rather than an evaluative stance. In this case, the LLM functioned as a second coder that deepened interpretative engagement rather than merely reinforcing human judgement.

In Stage 3, we used the Llama3 model as an additional coder to assign sentiment labels to manually extract student quotes. We developed Python scripts that automated model inference, eliminating the need for manual copying and pasting into MetaAI's interface and allowing student quotes to be read directly into structured prompt templates, thereby enabling rapid sentiment labeling, prompt iteration, immediate assessment of output relevance. This efficiency is valuable when open-ended responses can be lengthy and numerous. The model returned successful responses for all prompts submitted via its API, with no errors or interruptions encountered during inference.

Our quantitative sentiment labeling approach allowed us to examine data from 51 students across three time points, encompassing 535 quotations partitioned by seven student identities and lived experiences. As a rough estimate, the LLM processed approximately two to three quotes in the time it typically took the human researcher to label a single quote. The human labeling sessions were generally limited to one to two hours due to cognitive fatigue; however, the LLM processed batches of quotes continuously. Table 3 shows sentiment distributions across time points.

| **Table 3. Sentiment classification frequencies** | | | | |
|---|---|---|---|---|
| **Reflection time point** | **Positive sentiment counts** | **Negative sentiment counts** | **Neutral sentiment counts** | **Total sentiment counts** |
| Reflection 1 | 56 | 43 | 37 | 136 |
| Reflection 3 | 104 | 71 | 107 | 282 |
| Reflection 4 | 77 | 18 | 22 | 117 |

In Stage 4, we evaluated the extent to which LLM-generated sentiment labels aligned with those assigned by a human coder. Cohen's kappas for inter-rater reliability were calculated between the human coder and Llama3's output to assess the level of agreement on sentiment labels for verbal language and communication quotes extracted from students' reflections. The results indicated substantial for Reflection 1 ($\kappa = 0.52$), Reflection 3 ($\kappa = 0.64$), and Reflection 4 ($\kappa = 0.67$). These results suggest that, when used for well-scoped tasks such as sentiment labeling of pre-identified quotes, LLMs can produce outputs that are fairly consistent with human judgement. LLM-assisted sentiment analysis can support researchers in quickly characterizing affective trends in qualitative data, allowing subsequent thematic analysis to focus more deeply on meaning-making and interpretation rather than categorization. Similarly, LLM-assisted sentiment analysis could provide instructors with a high-level overview of sentiments in their classrooms.

#### 5.1.2 | Drawbacks: Post-processing burden and prompt sensitivity

Preliminary efforts to integrate the Llama3 model into our data analysis pipeline involved prompting it during Stage 2 to extract quotes mentioning students' verbal language and communication behaviors. We employed a similar iterative prompting approach for quote extraction as we did for sentiment labeling.

However, the model faced several challenges. One limiting factor was the input word limit for the Llama family of models, which necessitated splitting each student's reflection into multiple overlapping segments to maintain context. Despite this adjustment, the model struggled to yield a list of quotes relevant to students' verbal language and communication behaviors compared to those extracted by the human coder. As a result, we decided to rely solely on the manually extracted quotes for downstream tasks in our pipeline.

A primary challenge involved filtering and structuring the model's outputs to isolate relevant content. Even when the model generally adhered to prompt instructions, it frequently generated extraneous text, such as greetings and explanatory commentary, that required additional post-processing. The filtering burden was pronounced when the model produced multiple sentiment labels for a single quote or assigned sentiment labels to elements of the prompt itself, often defaulting to neutral categorizations. For example, in response to the student quote "The people are great, the train system makes sense, and the language barrier can be mostly solved with patience and google translate (Reflection 1, Cody)," the model generated the following output "I'm excited to dive into the sentiment analysis! Here's my output: 1. POSITIVE: student appreciates the people, train system, and ease of overcoming language barriers with patience and technology 2. NEUTRAL: student is taking time to reflect and gather their thoughts (no explicit positive or negative sentiment expressed) Additional comments: Overall, this journal entry seems to be a reflection on the student's positive experience in a foreign country. The text excerpts highlight the pleasant aspects of the environment, with no strong negative sentiments expressed." These behaviors contributed to discrepancies between human and model-generated labels and helped explain the lower strength of agreement observed by Reflection 1. Specifically, 20% of the sentiment labels generated by Llama3 for Reflection 1 were not identified by the human coder, compared to 8% for Reflection 3 and 7% for Reflection 4. As a result, human review remained necessary to ensure analytic coherence, partially offsetting the efficiency gains associated with automated labeling.

Another limitation is the need for extensive prompt refinement (Appendix A1). We engaged in iterative prompt tuning to account for the structure, length, and linguistic features of students' reflective writing. We highlight that prompt designs are often dataset- and context-specific, limiting their generalizability across research settings and may increase the upfront labor required before reliable outputs can be obtained.

In addition to analytic challenges, integrating LLMs into our workflow introduced substantial administrative overhead. Developing Python scripts to automate model inference, creating spreadsheets to track quotes and sentiment labels, and organizing outputs for inter-rater reliability analysis required considerable technical setup and data management. Although the software we used was open source and posed no access barriers, it entailed a learning curve and should be a consideration for research teams without prior coding experience. Although this infrastructure enabled more efficient processing once established, the upfront investment raised questions about its overall payoff, particularly for smaller datasets or analyses for single use.

### 5.2 | RQ2: Case study findings

Guided by Fig. 1c, we first share first the quantitative findings. Then, we present the qualitative findings and interpretation of connected results together to draw attention to longitudinal changes for all students as well as differences in changes for students with and without prior experience living abroad.

#### 5.2.1 | Quantitative findings

Statistically significant differences in positive and negative sentiment expression were observed between students with and without prior experience living abroad across all three reflection time points (Table 4). Students who had lived abroad (LA) expressed a greater number of negative sentiments about their verbal language and communication behaviors in Reflection 1 compared to students who had not lived abroad ($U$ = 318.5, $p$ < .05, $r_g$ = -.289). Students who have not lived abroad (not-LA) expressed a greater number of negative sentiments about their verbal language and communication behaviors in Reflection 3 compared to LA students ($U$ = 166.5, $p$ < .05, $r_g$ = .326). Not-LA students expressed a greater number of positive sentiments in Reflection 4 compared to LA students ($U$ = 152.5, $p$ < .05, $r_g$ = .383).

Results for other student identities and lived experiences without statistically significant differences are provided in Appendices 2–7.

**Table 4. Results of statistical tests comparing students with (n=13) and without (n=38) experience living abroad.**

| **Reflection time point** | **Sentiment** | **Group** | **Number of sentiments** | **Shapiro-Wilk *p* value** | **Mann-Whitney *U*** | **Mann-Whitney *p* value** | **Glass biserial correlation coefficient ($r_g$)** |
|---|---|---|---|---|---|---|---|
| Reflection 1 | Positive | LA | 13 | .008** | 276.5 | .511 | .119 |
| | | Not-LA | 43 | <.001*** | | | |
| | Negative | LA | 16 | .041* | 318.5 | .050* | -.289 |
| | | Not-LA | 27 | <.001*** | | | |
| | Neutral | LA | 8 | <.001*** | 266.5 | .650 | .079 |
| | | Not-LA | 29 | <.001*** | | | |
| Reflection 3 | Positive | LA | 24 | .051 | 270.5 | .609 | .095 |
| | | Not-LA | 80 | <.001*** | | | |
| | Negative | LA | 11 | .010** | 166.5 | .037* | .326 |
| | | Not-LA | 60 | <.001*** | | | |
| | Neutral | LA | 24 | .002** | 304.0 | .208 | .231 |
| | | Not-LA | 83 | <.001*** | | | |
| Reflection 4 | Positive | LA | 13 | .007** | 152.5 | .017* | .383 |
| | | Not-LA | 64 | .002** | | | |
| | Negative | LA | 3 | <.001*** | 283.5 | .318 | .148 |
| | | Not-LA | 15 | <.001*** | | | |
| | Neutral | LA | 3 | <.001*** | 298.0 | .182 | .206 |
| | | Not-LA | 19 | <.001*** | | | |
| All reflection | Positive | LA | 50 | .192 | 316.5 | .129 | .281 |
| | | Not-LA | 187 | .012* | | | |
| | Negative | LA | 30 | .103 | 284.0 | .423 | .150 |

| time points | | Not-LA | 102 | .040* | | | |
|---|---|---|---|---|---|---|---|
| | Neutral | LA | 35 | .028* | 310.5 | .169 | .257 |
| | | Not-LA | 131 | .019* | | | |
| Note: Statistical significance for both the Shapiro-Wilk test and Mann Whitney U test is denoted as $*p < .05$, $**p < .01$, and $***\ p < .001$. | | | | | | | |

### 5.2.2 | Qualitative findings and connected results

As mentioned in 4.4.3, reflections were characterized into four major themes. Table 5 shows each theme and representative quotes for positive and negative sentiments.

**Table 5. Major themes and representative quotations for positive and negative sentiments.**

| Theme | Sentiment expression | Representative student quotation |
|---|---|---|
| Prior experiences with languages and cultures beyond students' primary ones | Positive | "Since American pop culture and English are so pervasive, even in a homogenous, non-English speaking country like Japan, thanks to the globalizing market, I think my experience in America will help me communicate and relate to Japanese people discussing American culture." (Reflection 1, Ryan) |
| | Negative | "I felt mostly nervous prior to the trip. Having never really traveled out of the country, much less to non-English speaking locations, I was really nervous to manage making my way around Japan." (Reflection 1, Nicole) |
| Preparations for anticipated verbal interactions with Japanese locals | Positive | "Despite not being fluent in Japanese, I have been dedicating my free time to learning some of the languages. I know some of the basic expressions and sentences to get around and order food, however, I believe that Google Translate will be very helpful and key when trying to communicate with people when communication becomes difficult." (Reflection 1, Sebastian) |
| | Negative | "I guessed this when I was learning, but I was using a free course I found online and wasn't sure what else to do. I wish I had found some Japanese vlogs and watched those to at least potentially get a feel for a casual tone of conversation and keywords used." (Reflection 3, Morgan) |
| Language dynamics and their role in verbal interactions with Japanese locals | Positive | "In regard to locals who I may interact with at a store or other location where I need to interact for a purchase, I have had overall great experiences. It is sometimes difficult to communicate which is natural but I have used a lot of Google Translate which comes in handy and is able to make the experience a lot easier and both me and the Japanese person I am conversing with are more able to understand each other and can usually laugh off the |

| | | situation and understand each other pretty well to an extent." (Reflection 3, Erin) |
|---|---|---|
| | Negative | "This has gotten me by, but it isn't the best if I want to connect with the locals in Japan. The language barrier is very difficult, and it proves as an obstacle to having a fluid conversation like the ones I have in America." (Reflection 3, Gabriel) |
| Perceived outcomes of participating in study abroad | Positive | "I got really comfortable with the few words and phrases in my repertoire, and it was able to get me through most things throughout the day, of course it is not perfect and it can be frustrating sometimes, but I learned that it is not impossible." (Reflection 4, Adrian) |
| | Negative | "However, with this realization has also shown me how much more I need to go in order to be able to go "completely Japanese" while working in Japan in the future. For this reason, I only say I have grown slightly more comfortable speaking Japanese now because I am even more frustrated at myself when I make mistakes." (Reflection 4, Jack) |

In Reflection 1, LA students expressed more negative sentiments than Not-LA students. LA students acknowledged their experience with languages other than English but expressed sentiments suggesting a narrow view of how they could leverage their prior language and communication skills in Japan. Often, they acknowledged that they knew two or more languages but could not see how these skills might help them, as the languages were dissimilar to Japanese (e.g., Spanish, Chinese, Arabic). Not-LA students echoed similar sentiments, noting differences between Japanese and the languages they were familiar with, particularly in terms of language structure and alphabet. Although they did not explicitly describe how they might leverage their prior language knowledge, this group emphasized communication aids they intended to use, such as translation tools, to supplement their limited Japanese skills. They also viewed this barrier as motivation to learn basic Japanese phrases to facilitate communication with locals.

Although both groups acknowledged the language barrier at the beginning of the program, their approaches to navigating it varied. LA students expressed apprehension about verbally communicating with locals, largely due to concerns that their limited or nonexistent Japanese proficiency could lead to unintentionally offensive misunderstandings, sometimes leading them to withdraw from conversations if they found it difficult to convey their message. In contrast, Not-LA students discussed their communication perseverance, including leveraging the locals' familiarity with English (even if it was limited).

In Reflection 3, both groups of students reported ongoing challenges navigating the language barrier–especially engaging in meaningful conversations with locals–and how that impacted their experiences. Students noted differences in the quality of conversations compared to those they were accustomed to back home. Others felt relieved when locals were able to accommodate to English. Negative sentiments regarding expectations about language barriers were common in both groups. For LA students, negative sentiments were shaped by confirmation that their language barrier expectations were accurate. For Not-LA students, these feelings stemmed from expectations that navigating the language barrier would be easier than it ultimately proved to be, including expecting a higher prevalence of English in Japan. They also expressed frustration when their limited verbal skills could not be supplemented by other means of communication, such as written language.

Among Not-LA students, reflections highlighted how a lack of knowing how to prepare for language barriers limited their ability to engage in conversations or how their preparation proved unhelpful. For instance, some students focused on learning to read Japanese but realized that learning basic conversational phrases would have been more beneficial. Others expressed regret for not learning more Japanese to show locals that, as foreigners, they had attempted to understand the language. This group also tended to express low confidence in their ability to acquire skills needed to navigate the language barrier, sometimes comparing themselves to other multilingual students.

In Reflection 4, Not-LA students had positive sentiments centered on perceived improvements in their ability to navigate language barriers with individuals who may not speak their primary language. Many expressed increased confidence in their communication skills. For LA students, verbal communication takeaways from the study abroad experience included recognition of the importance of communication preparation before the program. Newfound confidence often stemmed from the friendliness and patience of Japanese locals towards students' attempts to communicate verbally, even when they were not fluent. This supportive environment encouraged students to treat verbal communication as a learning opportunity and motivated them to persist and work through communication challenges. Both groups recognized the use of communication aids, such as mobile translation apps, to supplement their communication efforts. However, the Not-LA group viewed these tools as an auxiliary tool, while the LA group described using them as a last resort solution when verbal communication was unsuccessful.

Figure 3 illustrates shared longitudinal sentiment expressions between both groups of students, while Figure 4 shows contrasting patterns. Statistically significant differences were observed between the two groups. LA students expressed more negative sentiments about their intercultural communication at the beginning of the program. They found it challenging to leverage their prior language and communication skills in Japan and felt apprehensive about verbally communicating with Japanese locals because of their limited Japanese proficiency. In comparison, Not-LA students reported more negative sentiments in the middle of the program but shifted to more positive sentiments by the end. Initially, the Not-LA group faced misaligned expectations regarding the language barrier, perceiving their preparation efforts as inadequate or unhelpful, and reported low confidence in their ability to develop Japanese verbal communication skills. Ultimately, the challenges the Not-LA group encountered with language and communication evolved into valuable learning opportunities that could be applied to future travels abroad.

| | Reflection 1 | Reflection 3 | Reflection 4 |
|---|---|---|---|
| Findings from students both **with and without** experience living abroad | Acknowledged language barrier and anticipated communication challenges<br><br>Expressed uncertainty about leveraging prior language experience (East Asian and non-East Asian) in Japan | Acknowledged ongoing challenges navigating language barrier to have meaningful conversations<br><br>Noted prior language experience (East Asian and non-East Asian) is not helpful | Expressed more comfort navigating language barriers<br><br>Acknowledged use of aids (e.g., translation apps) to supplement verbal communication efforts |

**Figure 3. Shared longitudinal trends in sentiment expression for students with and without prior experience living abroad.**

| | Reflection 1 | Reflection 3 | Reflection 4 |
|---|---|---|---|
| Findings predominantly from students **with** experience living abroad | Felt apprehensive about verbal communication with locals due to lack of Japanese fluency<br><br>Expressed concerns with language barriers leading to misunderstandings with locals | Reported that their preconceived concerns about language barriers were affirmed by their experiences | Framed communication aids as a last-resort solution when verbal communication was unsuccessful |
| Findings predominantly from students **without** experience living abroad | Mentioned perseverance in verbal communication efforts via communication aids (e.g., translation apps) | Recognized misaligned expectations about ease of navigating the language barrier<br><br>Expressed frustration with their unfamiliarity with alternative communication strategies (e.g., written communication) as substitutes for verbal interaction<br><br>Reflected on lack of preparedness and low confidence in developing necessary language skills | Recognized growth in navigating language barriers with others<br><br>Framed communication aids as an auxiliary tool to enhance communication experiences |

**Figure 4. Contrasting longitudinal trends in sentiment expression for students with and without prior experience living abroad.**

## 6 | Discussion and implications

### 6.1 | LLMs broaden variable exploration but introduce data workflow trade-offs

The Llama3 LLM's role as an additional coder supported our quantitative analysis by helping identify differences between demographic groups and longitudinal trends, insights that are time-consuming and challenging to detect through human-centered analysis alone. A few recent engineering study abroad studies focus on analyzing qualitative data with comparable sample sizes, but their findings consider only a few participant variables and do not provide longitudinal insights (Davis & Knight, 2025; Özkan et al., 2024). Sentiment analysis alone cannot replace qualitative analysis but can help pinpoint the most compelling starting points. Our human-LLM Cohen's kappa values (0.52, 0.64, 0.69) are similar to and slightly better than previously reported values for GPT-3.5 (κ=0.58) (Shaikh et al., 2023) and GPT-4 (κ=0.588) (Satya Putra et al., 2025) sentiment analysis in education research. Discrepancies between the human and LLM coder sentiment labels initiated deeper analytic discussion and encouraged refinement of coding decision.

Qualitative education research uses varied methodological approaches (Case & Light, 2011), but its defining characteristics of being time- and labor-intensive have been addressed through computer-aided methods since the late 1990s (Webb, 1999). Existing software (e.g., NVivo, Dedoose) has enhanced workflow efficiency by offering researchers user-friendly interfaces for managing data. LLMs with advanced language processing capabilities introduce new possibilities for qualitative analysis, raising questions about their capacity to mirror the nuanced methods of human analysis and whether their use streamlines or complicates existing workflows. Our integration of Llama3 required time-intensive iterative prompt engineering and manual post-processing. The reflective, journalistic style of the study abroad data posed challenges, as the writing sometimes referenced other entries, and extended context was not always preserved because of prompt length restrictions. Although we provided specific instructions to the Llama3 model (Appendix A1), it often included irrelevant text (e.g., statements such as "I'd be happy to help!"), necessitating human review. There were also instances where Llama3 assigned sentiment labels to the instructions listed in the prompt, provided redundant analysis for a single quote, and outputted an out-of-sequence list of sentiments, making it difficult to match quotes with corresponding Llama3 sentiment labels. These edge cases complicate automation for post-processing as their accuracy and reliability are not guaranteed. Transparency in how researchers develop prompts for different use cases may help advance prompt standardization, improve LLM outputs, and reduce time inefficiencies in key parts of the workflow. For example, we found that a low agreement for Reflection 1 came from Llama3's tendency to generate a more detailed analysis than requested, resulting in multiple sentiment labels for a single quote. There were also instances where the model assigned sentiment labels to the prompt instructions, often categorizing them as neutral. This was particularly evident in the higher proportion (20%) of Llama3 sentiment labels generated for Reflection 1 that the human researcher did not identify, in contrast to Reflection 3 (8%) and Reflection 4 (7%).

### 6.2 | Implications for LLM-assisted integrated computational and qualitative "mixed-methods" research

Although LLM capabilities continue to improve through training and fine-tuning for specialized tasks, their use remains limited by biases and opacity. Unlike researchers who contextualize data with subject knowledge, LLMs recognize language patterns and generate new language solely based on training data. Traditional practices like peer debriefing and data triangulation foster trustworthiness in qualitative work (Schwandt et al., 2007); however, these approaches depend on co-constructed ideas from the research team's collaborative efforts. The objectivity of language models has become an active area of research, particularly in applications where their usage has significant implications. Currently, however, there are no established standards for translating these practices into LLM-assisted qualitative analysis pipelines.

LLMs' generation of language without communicative intent is often compared to a stochastic parrot (Bender et al., 2021). Some propose simulating human reasoning via "self-correction," where language models refine their previous responses (Madaan et al., 2023; Welleck et al., 2022). However, this is a probabilistic process and prior research in LLM self-correction has shown that language models may struggle to improve responses, with performance sometimes deteriorating further (Huang et al., 2024). As a result, efforts may lead to increased disagreement between human coders and the model, rather than achieving resolution. It is essential to maintain a human-in-the-loop approach when integrating contributions from an LLM into qualitative data analysis. Researchers should be involved in all decision-making aspects of the analysis and interpret LLM outputs within the specific scope of the research topic. It is also crucial to consider the ethical implications of using LLMs, particularly how open source models offer significant ethical benefits and reduced risks compared to closed source models, extending beyond protecting participants' data (Eiras et al., 2024; Mathis et al., 2024). Open source communities have devoted significant efforts to addressing LLM inefficiencies that contribute to negative environmental impacts, including experimenting with hardware configurations to optimize the performance of smaller LLM architectures that require fewer resources to run.

Future research can focus on developing roadmaps for the ethical and meaningful integration of LLMs in qualitative research, considering their impact on various aspects of studies, from design and recruitment to data evaluation and reporting. Furthermore, a "two-heads-are-better-than-one" approach that strategically ensembling LLMs may offer improved task execution across qualitative data analysis. This method has proven effective in sentiment analysis applications, especially when paired with voting mechanisms that weight model outputs differently to reach a joint decision. Such an approach can emulate the corroborate process of integrating multiple perspectives, potentially adding robustness to the analysis (Delgadillo et al., 2024; Etelis et al., 2024).

### 6.3 | Living abroad impacts longitudinal sentiments of language and communication behaviors

In this case study, we explored the relationship between undergraduate students' identities and lived experiences and their sentiments regarding verbal language and communication behaviors across three time points in a month-long study abroad program. From the seven identities and lived experiences we explored, our results indicate that students' prior experience living abroad was the only factor that showed statistically significant differences in sentiment expression across all time points.

Prior work suggests that participating in study abroad can serve as a "cultural eye opener" (p. 1703) for students with low international experience, i.e., having visited fewer than five countries (Iskhakova et al., 2022). This can be attributed to students gaining knowledge and awareness during study abroad as they are exposed to new situations, resulting in foundational cultural insights that students with prior travel experience may have already acquired (Davis & Knight, 2025). Similarly, Mu et al. (2022) observed that networking with locals, immersing oneself in the host country's language, and stepping out of comfort zones serve as catalysts for students to recognize development of their intercultural communication skills. Our findings are congruent, revealing a longitudinal shift in students' sentiment expression toward their language and communication skills throughout the program, particularly among those who had no prior experience living abroad. Students who had never lived abroad initially expressed negative sentiments focused on limited views of how their existing language skills could apply in Japan, expressing frustration with their struggles to navigate language barriers and have meaningful conversations with locals. This group was also more negative midway through the program compared to their peers who had previously lived abroad, but their reflections became notably more positive by the end. In the last reflection, we observed an increase in students' confidence navigating language challenges, and their recollection of experiences with communication barriers were often reframed as opportunities for growth rather than setbacks. Students who had previously not lived abroad also expressed a desire to apply lessons learned from this experience to future travel.

**6.4 | Implications for engineering study abroad**

Our work has implications for the programmatic design choices educators and study abroad administrators make to ensure students obtain the intended benefits from their experiences and develop intercultural competence for their careers. Qualitative findings from our study suggest that students' perceptions about their language and communication experiences, such as anticipated language barriers, are formed early, likely because students are accepted into these programs months before travel. Opportunities to foster community-building before programs commence may help alleviate students' concerns about language barriers, such as meet-and-greet events with study abroad faculty and alumni. Combining these strategies may help mitigate the effects of STEM faculty members' lower support for cultural learning and challenging ethnocentrism in short-term study abroad programs (Niehaus & Wegener, 2019). To address different student needs, study abroad administrators can encourage students with prior lived abroad experience to reflect on how they previously navigated a new culture and how that might translate to a new culture, even if the languages are very different. For students who have not lived abroad, administrators can prepare students by emphasizing the importance of learning everyday phrases rather than solely vocabulary.

Further research could explore how students' sentiments evolve across a broader range of study abroad contexts, aiming to better understand how institutional attributes and program structures shape students' experiences with language and communication. Furthermore, our study centers on students' verbal behaviors, but nonverbal behaviors and written language are also key factors that influence students' communication experiences in study abroad programs (Berka et al., 2021; Chédru & Delhoume, 2023; Levine & Garland, 2015). This is particularly relevant in countries like Japan, where written language holds deep cultural significance (Gottlieb, 2005). Probing these research directions can also help identify characteristics of short-term study abroad programs that offer experiences comparable in quality to longer-duration programs, thereby reinforcing the value and legitimacy of shorter programs.

## 7 | Conclusion

Our exploration of integrating large language models (LLMs) into qualitative analysis offers promising avenues for future research. Although we found LLMs effective for certain tasks, such as sentiment labeling, important questions remain regarding how to ensure the reliability and trustworthiness of qualitative research when leveraging generative artificial intelligence tools. Perhaps most importantly, this study provides a workflow for how to leverage LLM-assisted computational and qualitative mixed-methods research to more rapidly and feasibly probe qualitative findings between participants from multiple demographic groups at once. Through our case study analysis at three distinct time points and a broad range of student identities and lived experiences, we uncovered significant variations in sentiment expression related to verbal communication influenced by students' prior experience living abroad. Students without prior experience living abroad navigated language barriers and redefined their expectations, often by reframing challenges with communication as valuable learning opportunities.

## Acknowledgements

This work was supported by the United States National Science Foundation grants EEC-2217741 and EEC-2107008.

## 8 | References

**Appendix**

**A1 | Characteristics used to design a prompt for sentiment labeling of student quotes.**

Note: All rows containing prompt excerpts were input into the model simultaneously.

| **Characteristic** | **Description** | **Prompt excerpt** |
|---|---|---|
| Large language model role | A brief description of the perspective or lens that the LLM should consider prior to executing the task. | "Act as an expert sentiment analyst specializing in analyzing texts and identifying sentiments expressed in those texts." |
| Background information | A brief description that provides the LLM with situational context prior to giving it instructions. For example, providing a description of the type of data the LLM will analyze. | "I have a collection of student journal entries that express POSITIVE, NEGATIVE, and NEUTRAL sentiments. For each journal entry, a list of text excerpts was extracted." |
| Task instructions | Detailed, step-by-step set of guidance statements of how the LLM should execute the desired task(s). | "I would like for you to read through the list of excerpts and detect the sentiment expressed for each excerpt in the list.<br><br>First, read through the list of excerpts given to you in the <text> tag.<br><br>Second, create an enumerated list where each item is the overall sentiment (POSITIVE, NEGATIVE, and NEUTRAL) that corresponds to the excerpt.<br><br>Next to the sentiment, provide a short summary of things discussed in the excerpt that expressed that sentiment." |
| Task clarification | Any additional clarification or guidance relating to the instructions. | "Note that the number of sentiments in your output list should be equal to the number of excerpts provided in the list between the <text> tag. |
| | | The order of the sentiments in your output should follow the order of the excerpts provided in the list between the <text> tag.<br><br>Output the enumerated list first, then in a new line you can provide any additional comments you may have.<br><br>It is essential that you do not make up things that are not in the excerpts when doing your analysis.” |
| Example input/output | An example of what the LLM should expect as input and the desired output. Statements should be as explicit as possible, such as to describe specific formatting requirements. | “For example, the sentiment for the first excerpt in the input list should be the first item in the output list.<br><br>For example, for the following instance from a student journal entry:<br>“1. I am so excited to practice my Japanese<br>2. I am still a bit worried about the language barrier”<br><br>your response should look like:<br>“1. POSITIVE: student is excited to practice their Japanese<br>2. NEGATIVE: student is worried about the language barrier” |
| Edge-case alternatives | Guidance on how the LLM should address edge cases (if known). | “If there is no text to analyze or the text only consists of "nan" or "None", please output NONE IDENTIFIED.” |
| Text for analysis | Use indicators (e.g., brackets, tags) to differentiate between the instructions and the text the LLM should analyze. | “Now that you know your instructions, here is the text for you to analyze: <text> {student quotation} </text>.<br><br>Take a moment to reflect and gather your thoughts. When you are ready, please begin your analysis." |

**A2 | Differences in sentiment expression between students who (n=20) and who do not (n=31) identify as Asian/Asian American.**

| Reflection time point | Sentiment | Group | Number of sentiments | Shapiro-Wilk *p* value | Mann-Whitney *U* | Mann-Whitney *p* value | Glass rank biserial correlation coefficient ($r_g$) |
|---|---|---|---|---|---|---|---|
| Reflection 1 | Positive | Asian | 23 | .003** | 306.5 | .952 | -.011 |
| | | Not Asian | 33 | <.001*** | | | |
| | Negative | Asian | 16 | <.001*** | 341.0 | .527 | .100 |
| | | Not Asian | 27 | <.001*** | | | |
| | Neutral | Asian | 19 | .003** | 234.5 | .110 | -.244 |
| | | Not Asian | 18 | <.001*** | | | |
| Reflection 3 | Positive | Asian | 50 | .121 | 231.5 | .122 | -.253 |
| | | Not Asian | 54 | .003** | | | |
| | Negative | Asian | 21 | <.001*** | 391.0 | .109 | .261 |
| | | Not Asian | 50 | .010* | | | |
| | Neutral | Asian | 52 | .011* | 249.0 | .229 | -.197 |
| | | Not Asian | 55 | .011* | | | |
| Reflection 4 | Positive | Asian | 24 | .006** | 386.5 | .125 | .247 |
| | | Not Asian | 53 | .006** | | | |
| | Negative | Asian | 8 | <.001*** | 315.5 | .902 | .018 |
| | | Not Asian | 10 | <.001*** | | | |
| | Neutral | Asian | 6 | <.001*** | 349.5 | .357 | .127 |
| | | Not Asian | 16 | <.001*** | | | |
| All reflection time points | Positive | Asian | 97 | .145 | 386.5 | .125 | -.106 |
| | | Not Asian | 140 | .041* | | | |
| | Negative | Asian | 45 | .101 | 368.5 | .256 | .189 |
| | | Not Asian | 87 | .038* | | | |
| | Neutral | Asian | 77 | .257 | 251.0 | .254 | -.190 |
| | | Not Asian | 89 | .040* | | | |

<u>Note:</u> Statistical significance for both the Shapiro-Wilk test and Mann Whitney U test is denoted as *$p < .05$, **$p < .01$, and *** $p < .001$.

**A3 | Differences in sentiment expression between students who (n=32) and who do not (n=19) know one or more foreign language.**

| Reflection time point | Sentiment | Group | Number of sentiments | Shapiro-Wilk *p* value | Mann-Whitney *U* or Student *t*-test *t* | Mann-Whitney or Student *t*-test *p* value | Glass rank biserial correlation coefficient ($r_g$) or Cohen's *d* |
|---|---|---|---|---|---|---|---|
| Reflection 1 | Positive | Foreign language | 38 | <.001*** | 263.0 | .408 | -.135 |
| | | No foreign language | 18 | .005** | | | |
| | Negative | Foreign language | 25 | <.001*** | 278.5 | .601 | -.084 |
| | | No foreign language | 14 | <.001*** | | | |
| | Neutral | Foreign language | 29 | <.001*** | 281.0 | .628 | -.076 |
| | | No foreign language | 12 | <.001*** | | | |
| Reflection 3 | Positive | Foreign language | 71 | .015* | 230.0 | .141 | -.243 |
| | | No foreign language | 33 | <.001*** | | | |
| | Negative | Foreign language | 39 | <.001*** | 365.5 | .220 | .202 |
| | | No foreign language | 32 | .074 | | | |
| | Neutral | Foreign language | 58 | .001** | 366.0 | .217 | .204 |
| | | No foreign language | 49 | .007** | | | |
| Reflection 4 | Positive | Foreign language | 43 | .001** | 368.5 | .192 | .212 |
| | | No foreign language | 34 | <.001*** | | | |
| | Negative | Foreign language | 12 | <.001*** | 297.5 | .881 | -.021 |
| | | No foreign language | 6 | <.001*** | | | |
| | Neutral | Foreign language | 11 | <.001*** | 356.0 | .220 | .171 |
| | | No foreign language | 11 | <.001*** | | | |
| All reflection time points | Positive† | Foreign language | 152 | .062 | -.458† | .650† | -.104† |
| | | No foreign language | 85 | .075 | | | |
| | Negative | Foreign language | 80 | .052 | 330.5 | .607 | .132 |
| | | No foreign language | 52 | .019* | | | |
| | Neutral | Foreign language | 94 | .083 | 349.0 | .381 | .148 |
| | | No foreign language | 72 | .009** | | | |

Note: Statistical significance for both the Shapiro-Wilk test, Mann Whitney U test, and Student *t*-test is denoted as $*p < .05$, $**p < .01$, and $*** p < .001$.
†Statistical test ran for this variable was a Student *t*-test and Cohen's *d* is reported.

**A4 | Differences in sentiment expression between students who (n=26) and who do not (n=25) have at least one foreign-born parent.**

| Reflection time point | Sentiment | Group | Number of sentiments | Shapiro-Wilk *p* value | Mann-Whitney *U* or Student *t*-test *t* | Mann-Whitney or Student *t*-test *p* value | Glass rank biserial correlation coefficient ($r_g$) or Cohen's *d* |
|---|---|---|---|---|---|---|---|
| Reflection 1 | Positive | Foreign parent(s) | 31 | .001** | 295.0 | .560 | -.092 |
| | | No foreign parent(s) | 21 | .002** | | | |
| | Negative | Foreign parent(s) | 22 | <.001*** | 338.0 | .800 | -.117 |
| | | No foreign parent(s) | 25 | <.001*** | | | |
| | Neutral | Foreign parent(s) | 22 | <.001*** | 273.0 | .283 | .015 |
| | | No foreign parent(s) | 15 | <.001*** | | | |
| Reflection 3 | Positive | Foreign parent(s) | 61 | .099 | 248.0 | .138 | -.237 |
| | | No foreign parent(s) | 43 | .005** | | | |
| | Negative | Foreign parent(s) | 34 | .001** | 344.5 | .712 | .060 |
| | | No foreign parent(s) | 37 | .010** | | | |
| | Neutral | Foreign parent(s) | 53 | .002** | 338.0 | .808 | .040 |
| | | No foreign parent(s) | 54 | .003** | | | |
| Reflection 4 | Positive | Foreign parent(s) | 33 | .001** | 397.0 | .158 | .222 |
| | | No foreign parent(s) | 44 | .033* | | | |
| | Negative | Foreign parent(s) | 10 | <.001*** | 324.5 | 1.0 | -.002 |
| | | No foreign parent(s) | 8 | <.001*** | | | |
| | Neutral | Foreign parent(s) | 8 | <.001*** | 382.0 | .193 | .175 |
| | | No foreign parent(s) | 14 | <.001*** | | | |
| All reflection time points | Positive† | Foreign parent(s) | 125 | .096 | -.579† | .565† | -.162† |
| | | No foreign parent(s) | 112 | .101 | | | |
| | Negative | Foreign parent(s) | 66 | .123 | 330.0 | .931 | .015 |
| | | No foreign parent(s) | 66 | .019* | | | |
| | Neutral | Foreign parent(s) | 83 | .121 | 329.5 | .939 | .014 |
| | | No foreign parent(s) | 83 | .020* | | | |

Note: Statistical significance for both the Shapiro-Wilk test, Mann Whitney U test, and Student *t*-test is denoted as $*p < .05$, $**p < .01$, and $***\ p < .001$.
†Statistical test ran for this variable was a Student *t*-test and Cohen's *d* is reported.

**A5 | Differences in sentiment expression between students who (n=14) and who do not (n=37) identify as low-income.**

| Reflection time point | Sentiment | Group | Number of sentiments | Shapiro-Wilk *p* value | Mann-Whitney *U* | Mann-Whitney *p* value | Glass rank biserial correlation coefficient ($r_g$) |
|---|---|---|---|---|---|---|---|
| Reflection 1 | Positive | Low income | 16 | .028* | 256.5 | .965 | -.010 |
| | | Not low income | 40 | <.001*** | | | |
| | Negative | Low income | 14 | .032* | 228.5 | .496 | -.118 |
| | | Not low income | 29 | <.001*** | | | |
| | Neutral | Low income | 8 | .002** | 270.0 | .806 | .043 |
| | | Not low income | 29 | <.001*** | | | |
| Reflection 3 | Positive | Low income | 29 | .047* | 251.5 | .879 | -.029 |
| | | Not low income | 75 | .004** | | | |
| | Negative | Low income | 19 | .096 | 256.5 | .965 | -.010 |
| | | Not low income | 52 | <.001*** | | | |
| | Neutral | Low income | 26 | .040* | 299.0 | .390 | .154 |
| | | Not low income | 81 | <.001*** | | | |
| Reflection 4 | Positive | Low income | 19 | .002** | 279.5 | .659 | .079 |
| | | Not low income | 58 | .005** | | | |
| | Negative | Low income | 6 | <.001*** | 230.5 | .449 | -.110 |
| | | Not low income | 12 | <.001*** | | | |
| | Neutral | Low income | 1 | <.001*** | 169.5 | .011* | .346 |
| | | Not low income | 21 | <.001*** | | | |
| All reflection time points | Positive | Low income | 64 | .818 | 267.5 | .863 | .033 |
| | | Not low income | 173 | .005** | | | |
| | Negative | Low income | 39 | .159 | 246.0 | .789 | -.050 |
| | | Not low income | 93 | .032* | | | |
| | Neutral | Low income | 35 | .177 | 321.5 | .186 | .241 |
| | | Not low income | 131 | .013* | | | |

Note: Statistical significance for both the Shapiro-Wilk test and Mann Whitney U test is denoted as $*p < .05$, $**p < .01$, and $***p < .001$.

**A6 | Differences in sentiment expression between students who (n=10) and who do not (n=41) identify as first-generation college attendees.**

<table>
<tr><th>Reflection time point</th><th>Sentiment</th><th>Group</th><th>Number of sentiments</th><th>Shapiro-Wilk p value</th><th>Mann-Whitney U</th><th>Mann-Whitney p value</th><th>Glass rank biserial correlation coefficient ($r_g$)</th></tr>
<tr><td rowspan="6">Reflection 1</td><td rowspan="2">Positive</td><td>First generation</td><td>15</td><td>.095</td><td rowspan="2">142.5</td><td rowspan="2">.123</td><td rowspan="2">-.305</td></tr>
<tr><td>Not first generation</td><td>41</td><td><.001***</td></tr>
<tr><td rowspan="2">Negative</td><td>First generation</td><td>10</td><td>.035*</td><td rowspan="2">173.0</td><td rowspan="2">.422</td><td rowspan="2">-.156</td></tr>
<tr><td>Not first generation</td><td>33</td><td><.001***</td></tr>
<tr><td rowspan="2">Neutral</td><td>First generation</td><td>2</td><td><.001***</td><td rowspan="2">130.0</td><td rowspan="2">.025*</td><td rowspan="2">.366</td></tr>
<tr><td>Not first generation</td><td>35</td><td><.001***</td></tr>
<tr><td rowspan="6">Reflection 3</td><td rowspan="2">Positive</td><td>First generation</td><td>18</td><td>.575</td><td rowspan="2">221.0</td><td rowspan="2">.705</td><td rowspan="2">.078</td></tr>
<tr><td>Not first generation</td><td>86</td><td><.001***</td></tr>
<tr><td rowspan="2">Negative</td><td>First generation</td><td>17</td><td>.016*</td><td rowspan="2">168.5</td><td rowspan="2">.378</td><td rowspan="2">-.178</td></tr>
<tr><td>Not first generation</td><td>54</td><td><.001***</td></tr>
<tr><td rowspan="2">Neutral</td><td>First generation</td><td>20</td><td>.210</td><td rowspan="2">221.5</td><td rowspan="2">.696</td><td rowspan="2">.081</td></tr>
<tr><td>Not first generation</td><td>87</td><td><.001***</td></tr>
<tr><td rowspan="6">Reflection 4</td><td rowspan="2">Positive</td><td>First generation</td><td>20</td><td>.045*</td><td rowspan="2">273.5</td><td rowspan="2">.046*</td><td rowspan="2">-.334</td></tr>
<tr><td>Not first generation</td><td>57</td><td>.001**</td></tr>
<tr><td rowspan="2">Negative</td><td>First generation</td><td>5</td><td>.002**</td><td rowspan="2">172.5</td><td rowspan="2">.330</td><td rowspan="2">-.159</td></tr>
<tr><td>Not first generation</td><td>13</td><td><.001***</td></tr>
<tr><td rowspan="2">Neutral</td><td>First generation</td><td>3</td><td><.001***</td><td rowspan="2">232.0</td><td rowspan="2">.442</td><td rowspan="2">.132</td></tr>
<tr><td>Not first generation</td><td>19</td><td><.001***</td></tr>
<tr><td rowspan="6">All reflection time points</td><td rowspan="2">Positive</td><td>First generation</td><td>53</td><td>.744</td><td rowspan="2">163.5</td><td rowspan="2">.322</td><td rowspan="2">-.202</td></tr>
<tr><td>Not first generation</td><td>184</td><td>.006**</td></tr>
<tr><td rowspan="2">Negative</td><td>First generation</td><td>32</td><td>.499</td><td rowspan="2">157.0</td><td rowspan="2">.253</td><td rowspan="2">-.234</td></tr>
<tr><td>Not first generation</td><td>100</td><td>.017*</td></tr>
<tr><td rowspan="2">Neutral</td><td>First generation</td><td>25</td><td>.129</td><td rowspan="2">258.0</td><td rowspan="2">.208</td><td rowspan="2">.259</td></tr>
<tr><td>Not first generation</td><td>141</td><td>.003**</td></tr>
<tr><td colspan="8">Note: Statistical significance for both the Shapiro-Wilk test and Mann Whitney U test is denoted as *$p < .05$, **$p < .01$, and *** $p < .001$.</td></tr>
</table>

**A7 | Differences in sentiment expression between students who (n=12) and who do not (n=39) identify as first-generation Americans.**

| Reflection time point | Sentiment | Group | Number of sentiments | Shapiro-Wilk *p* value | Mann-Whitney *U* | Mann-Whitney *p* value | Glass rank biserial correlation coefficient ($r_g$) |
|---|---|---|---|---|---|---|---|
| Reflection 1 | Positive | First-generation American | 17 | .133 | 172.5 | .155 | -.263 |
| | | Not first-generation American | 39 | <.001*** | | | |
| | Negative | First-generation American | 10 | .015* | 227.0 | .877 | -.030 |
| | | Not first-generation American | 33 | <.001*** | | | |
| | Neutral | First-generation American | 7 | <.001*** | 256.5 | .589 | .096 |
| | | Not first-generation American | 30 | <.001*** | | | |
| Reflection 3 | Positive | First-generation American | 17 | .109 | 303.5 | .115 | .297 |
| | | Not first-generation American | 87 | <.001*** | | | |
| | Negative | First-generation American | 15 | .012* | 256.0 | .622 | .094 |
| | | Not first-generation American | 56 | .001** | | | |
| | Neutral | First-generation American | 19 | .023* | 299.0 | .140 | .278 |
| | | Not first-generation American | 88 | <.001*** | | | |
| Reflection 4 | Positive | First-generation American | 20 | <.001*** | 190.0 | .312 | -.188 |
| | | Not first-generation American | 57 | .001** | | | |
| | Negative | First-generation American | 3 | <.001*** | 246.0 | .743 | .051 |
| | | Not first-generation American | 15 | <.001*** | | | |
| | Neutral | First-generation American | 1 | <.001*** | 160.5 | .024* | .314 |
| | | Not first-generation American | 21 | <.001*** | | | |
| All reflection time points | Positive | First-generation American | 54 | .640 | 247.0 | .777 | .056 |
| | | Not first-generation American | 183 | .025* | | | |
| | Negative | First-generation American | 28 | .280 | 252.0 | .693 | .077 |
| | | Not first-generation American | 104 | .026* | | | |
| | Neutral | First-generation American | 27 | .076 | 157.5 | .044* | .327 |
| | | Not first-generation American | 139 | .002** | | | |

Note: Statistical significance for both the Shapiro-Wilk test and Mann Whitney U test is denoted as *$p < .05$, **$p < .01$, and *** $p < .001$.

**A8 | Description of Figure 1.**

- Figure 1a: Convergent parallel design. Quantitative and qualitative are collected concurrently. Each dataset is analyzed separately using quantitative and qualitative methods, respectively. This is followed by the interpretation of connected results from both strands.
- Figure 1b: Explanatory sequential design. Quantitative data is collected and analyzed first. Quantitative results inform the subsequent qualitative data collection and analysis. This is followed by the interpretation of quantitative and qualitative results.
- Figure 1c. This study's quasi-convergent parallel design. Qualitative data collection is followed by the concurrent analysis of quantitative and qualitative data. This is followed by the interpretation of connected results from both strands.

**A9 | Text from Figure 2.**

- Figure 2: Data analysis pipeline
- Stage 1; Human researcher reviews N=151 student reflections from 3 time points (Reflection 1, Reflection 3, Reflection 4); Analytic memos
- Stage 2; Verbal language and communication quotes: Human researchers extracts quotations relevant to students' verbal language and communication behaviors; Reflection 1: 136 quotes, Reflection 3: 282 quotes; Reflection 4: 117 quotes
- Stage 3; Assign sentiment labels to quotes: Human researcher and Llama3 coder assign positive, negative, or neutral sentiment labels to quotations
- Stage 4; Review sentiment labels: Human researcher revises, if appropriate, their sentiment labels after reviewing Llama3 coder's labels

- Stage 5; Conduct thematic analysis: human researcher conducts thematic analysis; Conduct statistical tests: human researcher sets up data partitions and conducts statistical tests, human researcher identifies statistically significant relationships between student identities/lived experiences and sentiments
- Stage 6; Connect qualitative and quantitative findings: human researcher interprets qualitative findings considering quantitative results

**A10 | Text from Figure 3.**

- Findings from students both with and without experience living abroad
- Reflection 1: Acknowledged language barrier and anticipated communication challenges, expressed uncertainty about leveraging prior language experience (East Asian and non-East Asian) in Japan
- Reflection 3: Acknowledged ongoing challenges navigating language barrier to have meaningful conversations, noted prior language experience (East Asian and non-East Asian) is not helpful
- Reflection 4: Expressed more comfort navigating language barriers, acknowledged use of aids (e.g., translation apps) to supplement verbal communication efforts

**A11 | Text from Figure 4.**

- Findings predominantly from students **with** experience living abroad
    - Reflection 1
        - Felt apprehensive about verbal communication with locals due to lack of Japanese fluency

- Expressed concerns with language barriers leading to misunderstandings with locals
    - Reflection 3
        - Reported that their preconceived concerns about language barriers were affirmed by their experiences
    - Reflection 4
        - Framed communication aids as a last-resort solution when verbal communication was unsuccessful
- Findings predominantly from students **without** experience living abroad
    - Reflection 1
        - Mentioned perseverance in verbal communication efforts via communication aids (e.g., translation apps)
    - Reflection 3
        - Recognized misaligned expectations about ease of navigating the language barrier
        - Expressed frustration with their unfamiliarity with alternative communication strategies (e.g., written communication) as substitutes for verbal interaction
        - Reflected on lack of preparedness and low confidence in developing necessary language skills
    - Reflection 4
        - Recognized growth in navigating language barriers with others

- Framed communication aids as an auxiliary tool to enhance communication experiences